\documentclass[12pt]{article}

\pdfoutput=1

\usepackage{amsmath}
\usepackage{bbm}
\usepackage{multirow}

\usepackage{color} 

\usepackage{amssymb}
\usepackage{graphicx}
\newcommand{\Sec}[1]{Section~\ref{#1}}

\newcommand{\Tab}[1]{Table~\ref{#1}}

\newcommand{\Fig}[1]{Figure~\ref{#1}}

\newcommand{\Eq}[1]{Eq.~(\ref{#1})}

\newcommand{\XV}{\ensuremath{\mathcal{X}}}
\newcommand{\YV}{\ensuremath{\mathcal{Y}}}

\newcommand{\XS}{\ensuremath{\tilde{X}}}
\newcommand{\YS}{\ensuremath{\tilde{Y}}}

\newcommand{\be}{\begin{equation}}
\newcommand{\ee}{\end{equation}}
\newcommand{\bea}{\begin{eqnarray}}
\newcommand{\eea}{\end{eqnarray}}

\begin{document}
\baselineskip 0.6cm

\def\simgt{\mathrel{\lower2.5pt\vbox{\lineskip=0pt\baselineskip=0pt
           \hbox{$>$}\hbox{$\sim$}}}}
\def\simlt{\mathrel{\lower2.5pt\vbox{\lineskip=0pt\baselineskip=0pt
           \hbox{$<$}\hbox{$\sim$}}}}
\def\one{\relax{\rm 1\kern-.25em 1}}

\begin{titlepage}

\begin{flushright}
UMD-PP-011-018
\end{flushright}

\vskip 1.0cm

\begin{center}

{\Large \bf 
New Measurements with Stopped Particles at the LHC
}

\vskip 0.6cm

{\large
Peter W. Graham$^1$, Kiel Howe$^1$, Surjeet Rajendran$^{1,2}$, and Daniel Stolarski$^{2,3}$
}

\vskip 0.4cm

{\it $^1$ Stanford Institute for Theoretical Physics, Department of Physics, Stanford University,\\Stanford, CA 94305} \\
{\it $^2$ Department of Physics and Astronomy, Johns Hopkins University, 
         Baltimore, MD 21218} \\
{\it $^3$ Center for Fundamental Physics, Department of Physics, University of Maryland,\\College Park, MD 20742} \\

\vskip 0.8cm

\abstract{Metastable particles are common in many models of new physics at the TeV scale.  If charged or colored, a reasonable fraction of all such particles produced at the LHC will stop in the detectors and give observable out of time decays.  We discuss strategies for measuring the type of decay (two- vs three-body), the types of particles produced, and the angular distribution of the produced particles using the LHC detectors.  We demonstrate that with a plausible level of control over experimental uncertainties and $\mathcal{O}(10-100)$ observed decay events, the gauge properties and some couplings of the new particles can be measured. If the new particle has a dominant three-body decay, then the spin properties of the particles and Lorentz structure of the decay operator can also be distinguished or constrained.  These measurements can not only reveal the correct model of new physics at the TeV scale, but also give information on physics giving rise to the decay at energy scales far above those the LHC can probe directly.}

\end{center}
\end{titlepage}


\tableofcontents

\section{Introduction}
\label{sec:intro}

The discovery of new long lived particles at the Large Hadron Collider (LHC) would be tremendously interesting and could shed light on the dark matter puzzle as well as on physics of very high scales.  Particles that are long lived on collider time scales ($\gg$ ns) which are colored or electrically charged will be slowed down by electromagnetic interactions in the detectors at the LHC, causing a fraction of them to stop~\cite{Drees:1990yw, Arvanitaki:2005nq}.  Observations of the eventual out-of-time decay of these stopped particles can, in many cases, unmask their gauge quantum numbers, spin, and the nature of the physics responsible for decay. In this paper, we present strategies to make measurements on the decays of stopped particles in the LHC detectors. Keeping in mind possible limitations of the detectors, we evaluate the prospects for identifying TeV and UV physics that couples these new particles to the Standard Model. 

Massive metastable colored or electrically charged particles (MMCP, referred to in the draft as $X$) generically arise if the particle is protected from rapid decay due to accidental symmetries of the low energy Lagrangian that are nevertheless violated by physics in the ultraviolet. The possible decay of the proton caused by the presence of GUT scale interactions that violate the accidental baryon number symmetry of the standard model is a well known example of this phenomenon. Metastable particles that carry color (SIMPs) or electric charge (CHAMPs) emerge naturally in several scenarios of physics beyond the standard model (for general reviews see refs. \cite{Fairbairn:2006gg, Raklev:2009mg}).  An example of this scenario is a charged next to lightest supersymmetric particle (NLSP) such as a stau or stop decaying via higher dimension operators to a gravitino or axino.  Other models with long lived charged particles include~\cite{Senjanovic:1984rw, Bagger:1984rk, Banks:1986cg,  Dimopoulos:1996vz, Ambrosanio:1997rv, Hinchliffe:1998ys, Frampton:1997up, Polonsky:2000zt,  Covi:2001nw, Goldberger:2002pc, Nomura:2003qb, Cheung:2003um,  Feng:2003nr,  Nomura:2004is, Nomura:2004it, ArkaniHamed:2004fb,  Chacko:2008cb, Arvanitaki:2008hq,  Walker:2009ei, Walker:2009en, Graham:2009gy, McDonald:2009ab, Craig:2011ev}. While metastability due to high scale physics is well motivated, the new particle may be long lived because of very small marginal couplings, such as in $R$-parity violating SUSY~\cite{Barbier:2004ez}, pure Dirac neutrinos~\cite{Gupta:2007ui}, or TeV scale See-Saw\footnote{A particle may also be long lived due to kinematics, namely $X$ is nearly degenerate with the final state it can decay to, but we will not consider this case because it is experimentally extremely difficult.}~\cite{Cheung:2011ph}.  

If such a new particle is discovered at the LHC, it will be crucial to directly measure as many of its properties as possible in order to determine the underlying physics models. It has been shown that using production and propagation, it is possible to measure the MMCP's mass~\cite{Hinchliffe:1998ys,Allanach:2001sd,Kilian:2004uj}, spin~\cite{Rajaraman:2007ae,Kitano:2008sa,Ito:2010xj,Buckley:2010fj}, color representation~\cite{Buckley:2010fj}, flavor content~\cite{Kitano:2008en,Feng:2009bd}, polarization~\cite{Kitano:2010tt}, as well as its coupling to the Higgs and other standard model particles~\cite{Chang:2011jk, Luty:2011hi}. It has also been shown that MMCP's can be used to measure other properties of the new physics sector~\cite{Choudhury:2008gb, Biswas:2009zp, Biswas:2009rba, Feng:2009yq,Ito:2009xy, Biswas:2010cd,Ito:2010un}.  There have been searches for slow-moving MMCP's at LEP~\cite{Heister:2002hp,Heister:2003hc,LEPWG}, the Tevatron~\cite{Abazov:2008qu,Aaltonen:2009kea,Abazov:2011pf}, and the LHC~\cite{Chatrchyan:2012,Khachatryan:2011ts,Aad:2011yf,Aad:2011hz,Khachatryan:2011cd} which place bounds on their production.  

While much can be learned by studying the production and propagation of $X$ particles at the LHC, there is interesting physics that can only be learned by studying their decays. For example, the Lorentz structure of the decay and the branching fractions to different standard model particles can constrain the UV physics causing the decay. Previous proposals to study decays of $X$s include looking for decays from the surrounding rock~\cite{De Roeck:2005bw}, and building a new detector to capture $X$ particles~\cite{Hamaguchi:2004df,Feng:2004yi,Hamaguchi:2006vu}. While these proposals could be implemented in the far future, it is interesting to see what measurements can be made with the detectors that are already in place. It is often the case that many $X$ particles will be stopped in the detectors~\cite{Arvanitaki:2005nq}, and the present detectors can sometimes be better at capturing MMCPs than potential new detectors~\cite{Feng:2004yi}. Observing decays within the detector can be quite challenging because the LHC apparatus were designed to measure particles coming from a central interaction point, while in our case the event originates elsewhere.  Despite these difficulties, both the D0~\cite{Abazov:2007ht} and CMS~\cite{Khachatryan:2010uf, Khachatryan:2011ab} have performed searches for decays of stopped particles, demonstrating that difficult experimental issues such as triggering can be solved.

Previous studies of decays have shown that it is in principle possible to measure the scale suppressing the operator which mediates decays as well as the spins of the $X$ in the context of certain models without taking into account experimental realities~\cite{Buchmuller:2004rq, Buchmuller:2004tm}.  More realistic studies have demonstrated how to measure the lifetime~\cite{Asai:2009ka, Pinfold:2010aq}, and how to measure the origin of longevity in the specific case of a stau NLSP~\cite{Ito:2011xs}.  In this paper we will show that topologies and kinematic distributions of out-of-time decays can realistically be measured at the LHC, and these measurements can reveal the properties of TeV-sector particles as well as test and motivate models of UV physics responsible for the decay.

In this context, the primary purpose of this paper is to provide theoretical guidance for the types of measurement and accuracy which could feasibly distinguish different scenarios of long-lived  particles based on observations of their out-of-time decays in the LHC detectors.  In Section \ref{sec:observation} we motivate and describe a simple model parameterizing the observables and uncertainties for experimental measurements of the details of out-of-time decays. We apply this in Section~\ref{sec:topologies} to determine the ability of the LHC to distinguish MMCP scenarios based on the decay products in several motivated scenarios and in general. In Section~\ref{sec:effectivetheory} we focus on MMCPs with three-body decays and determine to what degree the LHC can distinguish models with particles of different spin and interactions of different Lorentz structure. We then conclude.

\section{Observing Decay Properties at the LHC}
\label{sec:observation}

As will be discussed, the vast majority of stopped MMCPs will come to rest in the barrel calorimeters at ATLAS and CMS \cite{Arvanitaki:2005nq}. Therefore searches for late decays at ATLAS and CMS \cite{Khachatryan:2010uf} look for energy deposits in the calorimeters occurring out of time with beam crossings, which will be observable for MMCP lifetimes of $100{\rm~ns} \lesssim \tau \lesssim 1{\rm~year}$. In this section we describe a simplified model of the observables in out-of-time decays occuring in the LHC calorimeters, including the sources and treatment of systematic uncertainties, the expected event rates for a variety of MMCPs, and the reduction of cosmic ray backgrounds. For concreteness we work in the context of a specific search strategy focused on out-of-time decays originating in the electromagnetic calorimeter (ECAL) of CMS or ATLAS, but we note that most of our results in the following sections can be applied more generally to evaluate a variety of possible search strategies for observing the details of out-of-time decays in the LHC detectors and beyond.

\subsection{Observables and systematic uncertainties}

\begin{figure}
\centering
\includegraphics[width=16cm]{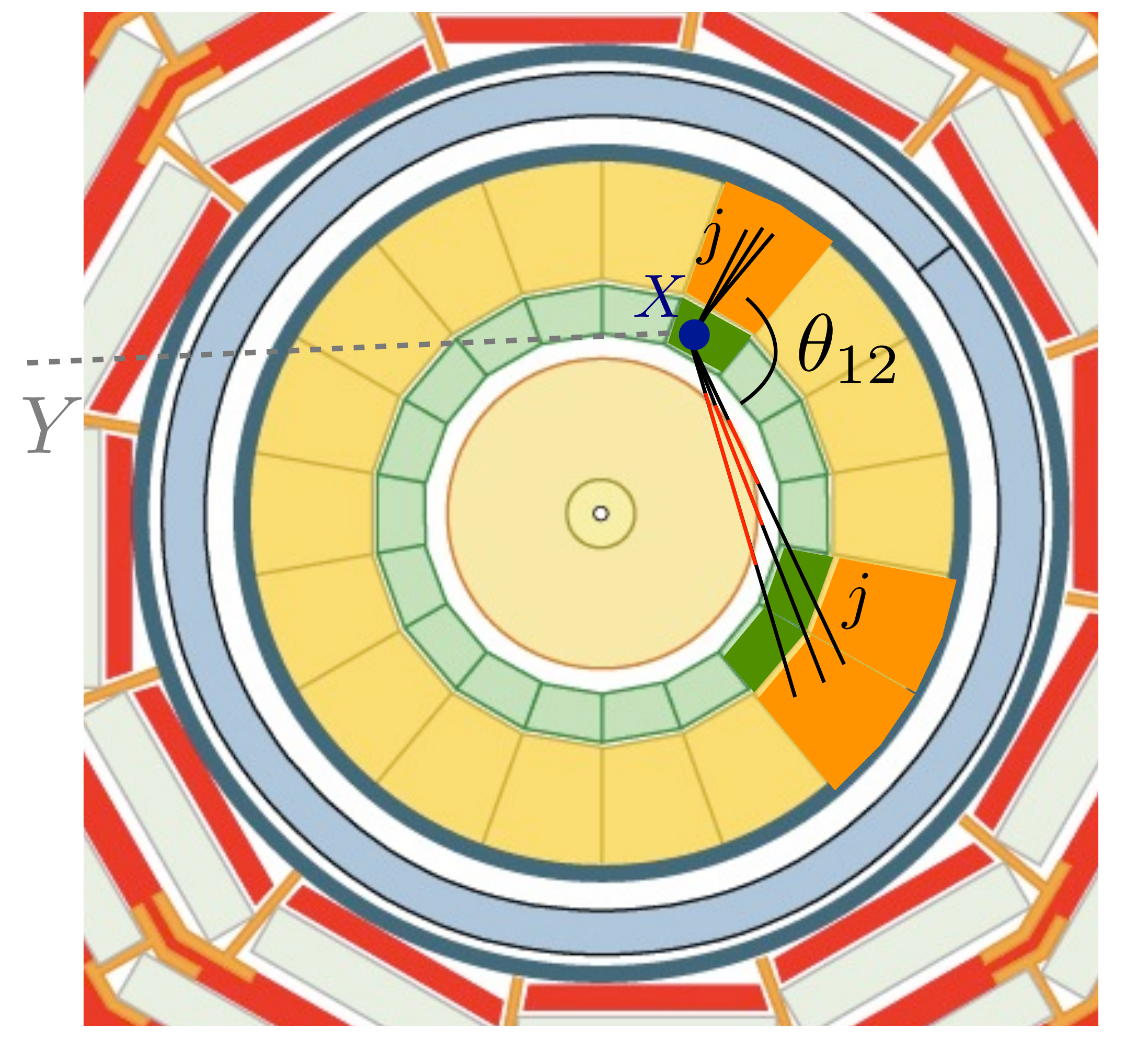}
\caption{\label{fig:strategy}  
A schematic view of the event geometry we are considering. This is a cross sectional $x-y$ view of a detector taken from~\cite{CMSDraw}.  The inner circle is the tracking chamber, the next (green) section is the electromagnetic calorimeter, followed by the (orange) hadronic calorimeter. The outermost layers are the muon chambers. An MMCP $X$ that had stopped in the ECAL decays into two jets and an invisible $Y$ which escapes the detector without interacting. The ECAL cell where the decay took place shows energy deposition.  One of the jets exits the ECAL and deposits its energy into the HCAL. The second jet exits the ECAL in another direction, leaving tracks which do not point back to the interaction region, and then depositing energy in a different region of both the ECAL and HCAL. This is a schematic representation; the $\phi$ resolution of the actual detector is much finer in both the ECAL and HCAL, so that the cells that lit up in the calorimeter should give enough information to approximately determine $\theta_{12}$, the angle between the jets.}
\end{figure}

For out-of-time decays originating in the barrel calorimeter as depicted in Figure~\ref{fig:strategy}, the ability to identify and measure the kinematic properties of final-state particles is very diminished by the geometry and stopping power of the calorimeters and the absence of information from the inner detector. There are however a variety of ways in which these obstacles could be overcome by exploiting the spatial segmentation of the calorimeters. To illustrate this point, we propose to look for events originating in the ECAL with a large fraction of the radiation escaping the ECAL and interacting with other parts of the detector, especially the hadronic calorimeter (HCAL). The motivations and details of this strategy are discussed in more detail in Appendix~\ref{app:measurement}. We find that in these types of events, the location of the decay vertex and the multiplicity and direction (though not energy) of jets and hard muons can then be reconstructed from the pattern of energy deposition in the finely segmented $\eta\times\phi$ plane of the barrel ECAL and HCAL, as illustrated in Figure~\ref{fig:strategy}. 

To understand the detailed response of the detectors to out-of-time off-vertex decays would require an event-by-event detector-level simulation. This is however beyond the scope of our analysis and unnecessary to demonstrate the observable features which will distinguish MMCP models and to explore the rough capabilities of the LHC detectors. Instead, we will work with events at the parton level and parametrize the experimental uncertainty with a simple model. 
After the initial hard decay of the MMCP and subsequent prompt decays of daughter particles, colored particles passing an energy cut $E_j$ are identified as hard ``jets", and muons passing an energy cut $E_\mu > 20{\rm~GeV}$ are identified as hard muons. To parametrize the uncertainty in detector response, we define an angular uncertainty $\Delta\theta$.  The measurement of the direction of each jet and muon in each decay event is taken from a gaussian distribution of full width $\Delta\theta$ around the true angle. As described in Appendix~\ref{app:measurement}, the geometry of the calorimeters motivates optimistic, nominal, and conservative benchmark values of respectively $\Delta\theta = 10^\circ, 30^\circ,{\rm~and~} 60^\circ$. $\Delta\theta$ also sets the angular separation at which a jet will be considered distinct in the analysis. Any colored particles within this separation are grouped as a single ``jet", and we restrict our attention also to muons isolated by $\Delta\theta$ from any jet. For a typical jet we will consider, the values of $\Delta\theta$ are greater than or comparable to the angular spreading of the jet due to soft colored radiation, so we ignore this effect.

Clearly the actual uncertainty in angular measurement will be non-isotropic and a function of the location of the decay vertex and the direction and energy of each object. Our approximation is partially justified by the fact that the locations of decay vertices in the calorimeters are uniformly distributed in the azimuthal direction, and that the overall orientation of the decays are isotropically distributed. Moreover, our results can be viewed as approximate targets for the degree to which the uncertainties in direction measurements must be understood in a full search to meet certain physics goals.

It is also important to consider how the triggers, energy cuts, and true uncertainty in direction measurement can distort the statistical distributions of event observables. In our simplified model, we will estimate the magnitude of these effects by studying the changes in these distributions as we adjust the parameters $\Delta\theta$ and $E_j$.

In Appendix~\ref{app:measurement} we discuss further the details, motivation, and expected performance at CMS and ATLAS for this strategy. However, in the analysis of Sections \ref{sec:topologies} and \ref{sec:effectivetheory}, we rely only on the assumption that the direction and multiplicity of muons and jets are measurable in out-of-time decays. The specific details of the measurement strategy do enter our results when we estimate signal rates by considering the stopping rates in the barrel ECAL, but limited explorations suggest that other strategies would obtain similar or worse efficiencies.

\subsection{Signal rates}
\label{subsec:signal}

To interpret our results it will be useful to make an estimate of the minimal event rates for a variety of MMCPs, taking into account the production cross sections and stopping rates of the MMCP and the trigger and cut efficiencies for out-of-time decays originating in the ECAL.

Electrically charged MMCPs are slowed down due to electromagnetic interactions with the detector material, and colored MMCPs will slow in the same way if they hadronize to charged states. A fraction $\delta$ of produced X will lose enough energy to stop within the barrel ECAL. We focus on the barrel ECAL because heavy $X$ that are directly pair produced will tend to be produced centrally and stop in the barrel calorimeters, although for more general models it may not be the optimal strategy.

To determine the 14 TeV LHC reach quantitatively, we take as a benchmark a hypothetical dataset of $200{\rm~fb}^{-1}$.  If a late decaying particle with lifetime $\gtrsim 100 \;{\rm ns}$ is discovered, the proton bunch structure could in principle be optimized to allow the decays of $\sim50\%$ of stopped MMCPs to be observed sufficiently out of time with any LHC collisions. Taking a further $50\%$ trigger/cut efficiency for late decay searches \cite{Khachatryan:2010uf}, in such a dataset the number of late decays observed in the ECAL for a particle with pair production cross section $\sigma$ will be
\begin{equation}\label{eq:eventrate}
 N = 25 \times \left(\frac{\sigma}{100\,{\rm fb}}\right)\left(\frac{\delta}{0.25\%}\right).
\end{equation}

For an MMCP of given spin and gauge representations, there is a minimal production rate due to direct production. Table~\ref{tab:eventrate} shows the corresponding mass at which $10$ and $100$ out-of-time decays with vertex in the ECAL would be observed in the benchmark dataset. As can be seen, statistically significant distributions of the experimental observables can be obtained for a variety of MMCPs over interesting mass ranges, even with the conservative assumption that model-dependent effects do not increase the overall production rate. The calculation of  stopping fraction $\delta$ is described in Appendix~\ref{app:stopping}.

\begin{table}
\begin{center}
 \begin{tabular}{c||c|c|c|c}
 & N & $M$ & $\sigma(M)$ & $\delta(M)$ \\
\hline
\multirow{2}{6cm}{Q=0 color octet fermion\\(gluino)} & 10 &  $\sim1300$ GeV & 40 fb & 0.25\%  \\ 
		 & 100 &  $\sim1000$ GeV & 400 fb &  0.25\% \\ 
\hline
\multirow{2}{6cm}{Q=2/3 color triplet scalar\\(stop)} & 10 & $\sim600$ GeV & 50 fb & 0.2\%  \\ 
		 & 100 &  $\sim400$ GeV & 400 fb & 0.25\% \\ 
\hline
\multirow{2}{6cm}{Q=1 SU(2) doublet fermion\\(higgsino)} & 10 & $\sim500$ GeV & 33 fb & 0.3\%  \\ 
		 & 100 & $\sim300$ GeV & 290 fb &  0.35\% \\ 
\hline
\end{tabular}
\caption{\label{tab:eventrate} The masses at which $N=10$ and $N=100$ observable late decay events are produced in the ATLAS or CMS ECAL with 200 fb$^{-1}$ at the 14 TeV LHC, assuming only direct production as in Eq.~(\ref{eq:eventrate}). The relevant direct production cross section $\sigma(M)$ and stopping fraction $\delta(M)$ calculated at each mass are shown. For colored particles, only colored production was included. Tree level cross sections and velocity distributions calculated with MadGraph 5 \cite{Alwall:2011uj}, NLO cross sections for gluino from \cite{Hewett:2004nw}.}
\end{center}
\end{table}

To get a feel for the numbers, consider a 1 TeV gluino.  At the 14 TeV LHC with our benchmark set of 200 fb$^{-1}$, 120,000 gluinos will be produced,  and we can use Eq.~(\ref{eq:eventrate}) and the calculation of $\delta$ in Appendix~\ref{app:stopping} to find that there will be about 75 observed decays in the ECAL.  As we will see in Sections~\ref{sec:topologies}~and~\ref{sec:effectivetheory}, this is enough to distinguish Split SUSY\footnote{Our benchmark models, including Split SUSY, will be described in Section~\ref{sec:models}.} from many other possible scenarios. Most other MMCP candidates also have large enough direct production rates at the LHC for our methods to be applicable.  However, for direct production of a right handed stau more than $\sim10$ decays in the ECAL will only be observed for a mass lighter than $90$ GeV. This is excluded by LEP searches for long lived charged particles \cite{Abbiendi:2003yd}, and therefore our methods are only applicable to a stau MMCP when the total LHC production cross section is enhanced by cascades to the NLSP.\footnote{Since this work originally appeared, stronger limits have been released \cite{Chatrchyan:2012}.  We discuss these in the conclusions.}

\subsection{Background rates}

The primary advantages of studying decays occuring out-of-time from beam crossings are that there is no stastistical background from competing processes in collisions, and no event-by-event background from pile-up or the underlying event. However, for MMCPs produced at low rates, cosmic rays become an important source of statistical background.

Ref. \cite{ATLAS:conf2010} describes the cuts used to reduce the cosmic background in ATLAS out-of-time decay searches, which all together reduce the cosmic sample by factor of $\sim 5\times10^5$. For a triggering rate in the calorimeters of $2~{\rm Hz}$ \cite{Boonekamp:2004}, this corresponds to a reduced background rate of $\sim20$ cosmic ray events per year in our benchmark dataset (with the reasonable assumption $5\times10^6$s of empty bunch crossings during live beam per year). Focusing on events with mid-ranged electromagnetic energy fractions, $0.1 < {\rm Jet~EMF} < 0.9$ as would be expected for a decay orginating in the ECAL, would reduce the rate to $\sim 1-5$ events per year \cite{ATLAS:conf2010}. This study included a cut $E >{\rm 50~GeV}$ for the leading jet, and the background rate could be easily reduced to $\mathcal{O}(1/{\rm yr})$ by increasing the cut while remaining sensitive to higher mass MMCP decays.

However, the cuts described in Ref. \cite{ATLAS:conf2010} include a cut on events with muon segments present which reduces the background by a factor of $\sim 1000$.  If we wish to measure muons in late decay events, the cosmic muon background needs to be reduced in another way. For example, because the cosmic spectrum falls off with energy, increasing the calorimeter energy selection to $E\gtrsim500$ GeV would reduce the background by 3 orders of magnitude based on the measured spectrum \cite{ATLAS:cosmics2011}. In most of the decays we consider the muons are relatively soft compared to the total energy of the event, and therefore a cut on muons with high energies relative to the calorimeter deposit could also be used to reduce the cosmic background while preserving the signal. Furthermore, because the cosmic muon background vanishes below the horizon, it is likely also possible to use track direction and timing measurements to veto the cosmic backgrounds while retaining a large fraction of signal events. This would lead to a reduction in the number of accepted signal events when muons are present, but because the overall decay orientation will be isotropically distributed, the total rate could in principle be inferred from the below-horizon rate. It seems plausible that the background can be reduced in these ways, but a more detailed analysis is beyond the scope of this paper and we do not include the effects of such cuts in our analysis. 

We therefore expect that the cosmic ray contamination of our benchmark dataset can be reduced to O(1). For a dataset of 100 signal events, this is comparable to both the systematic errors we will estimate from signal acceptance and the statistical uncertainties.

We note that Ref. \cite{ATLAS:conf2010} also employs cuts on jet shape which could in principle distort the distributions of our observables. However, these cuts are conservative and consistent with the shapes of energy depositions we expect from out-of-time decays involving large hadronic energy components. We model these effects only through our energy cuts on the parton-level ``jets" and muons.

At the high luminosities we are considering, beam related backgrounds could become important. These are however more easily reduced from the characteristic tracks and patterns of energy deposition \cite{ATLAS:conf2010}.

\section{Distinguishing Decay Topologies}
\label{sec:topologies}

In general a heavy MMCP will decay dominantly to low multiplicity primary states including on-shell heavy standard model particles. In many models a conserved quantum number (e.g. $R$-parity) requires the MMCP primary decay to also include another new particle associated with dark matter, which may be the dark matter candidate itself (e.g. neutralino, gravitino, axino) or directly connected to the dark matter sector (e.g. chargino).  We will refer to these as WIMPs/WIMP sector particles though our meaning is more generic than the normal usage. Measuring the properties and branching fractions for the primary products of an MMCP decay can in principle directly reveal properties of the MMCP, the WIMP sector, and the physics mediating the MMCP decay. However, in realistic scenarios this is obscured by the secondary decays of the heavy primary SM particles ($W$, $Z$, $h$, $t$), the possible secondary decay chain of the WIMP sector particle to the WIMP,  and the overall resolution and capabilities of the detector. In this section we discuss how these difficulties can be overcome, and how the late decay observables described in \Sec{sec:observation} can be used to distinguish models with sufficiently different decay modes.

\subsection{Benchmark Models}
\label{sec:models}
 
To explore the possibility of identifying MMCP properties and distinguishing different models, we consider several motivated benchmark cases with parameters chosen to give similar LHC event rates and signatures, summarized in the caption of Table~\ref{tab:topologies}.  Our objective  is not to explore all possible models and their parameter spaces, but rather to demonstrate that the observables we have identified provide interesting information about and can distinguish between a variety of motivated models.

\begin{figure}
\centering
\includegraphics[width=16cm]{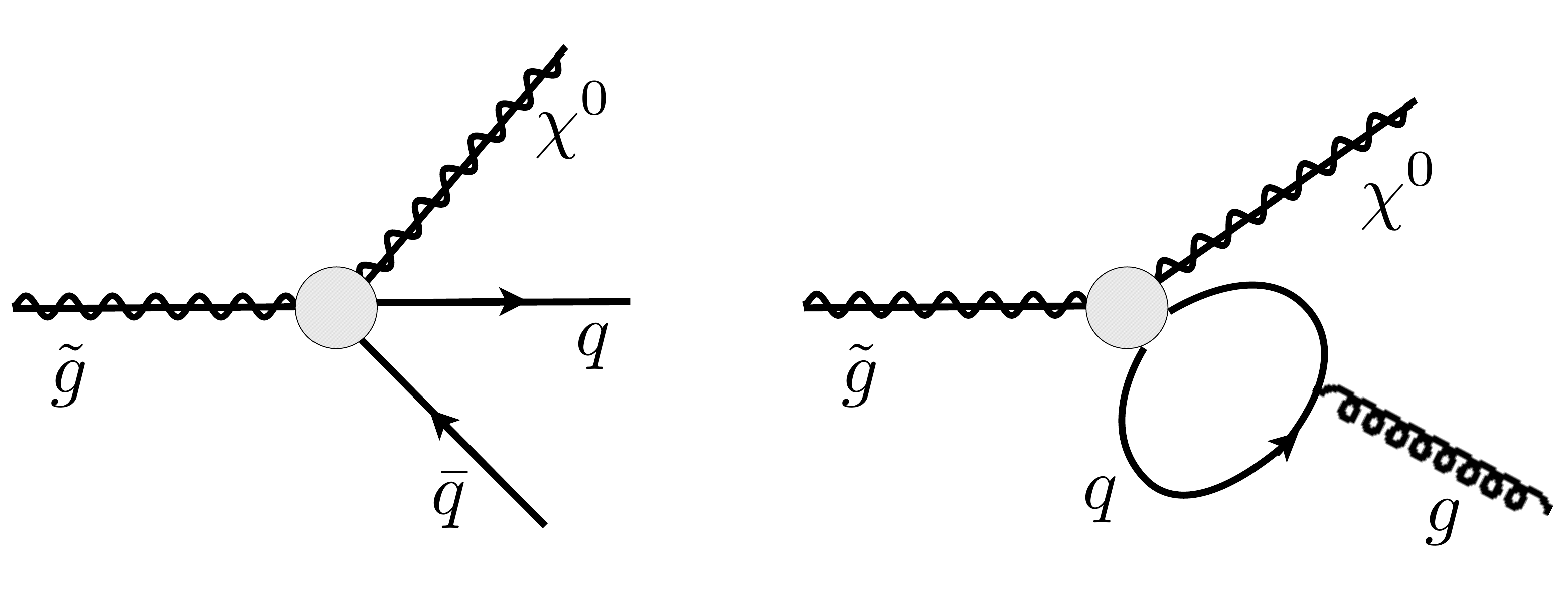}
\caption{\label{fig:GluinoFeyn}  
Feynman diagrams for the decays of a long lived gluino in Split SUSY.  The blob represents the four point vertex generated after integrating out the heavy squark.  The diagram on the left is a tree level three-body decay to two quarks and a neutralino, while the diagram on the right is the loop level two-body decay to a neutralino and a gluon. In certain regions of parameter space, the loop induced decay can become dominant over the three-body decay~\cite{Gambino:2005}. }
\end{figure}

In Split Supersymmetry~\cite{ArkaniHamed:2004fb}, the gluino is long lived because the squarks that mediate its decay are much heavier than 1 TeV. In particular, a 1-2 TeV gluino can have out-of time decays for scalar masses $10^7{\rm~GeV}\lesssim m_0 \lesssim 10^{11}{\rm~GeV}$ \cite{Gambino:2005}. Depending on the gluino and scalar mass scales, either a three-body decay to quarks and neutralino, or a two-body decay to quarks and neutralino/chargino can dominate, or both can compete \cite{Gambino:2005} -- we therefore consider the two-body and three-body decays separately. These two decays are shown in Figure~\ref{fig:GluinoFeyn}. To obtain the correct relic density for a weak scale WIMP, the Split SUSY neutralino is constrained to have a significant higgsino component \cite{Giudice:2004tc}. To explore the remaining freedom in the Split SUSY low energy parameter space, we consider two representative benchmark points motivated by gaugino mass unification boundary conditions and dark matter relic density.
 
\begin{figure}
\centering
\includegraphics[width=17cm]{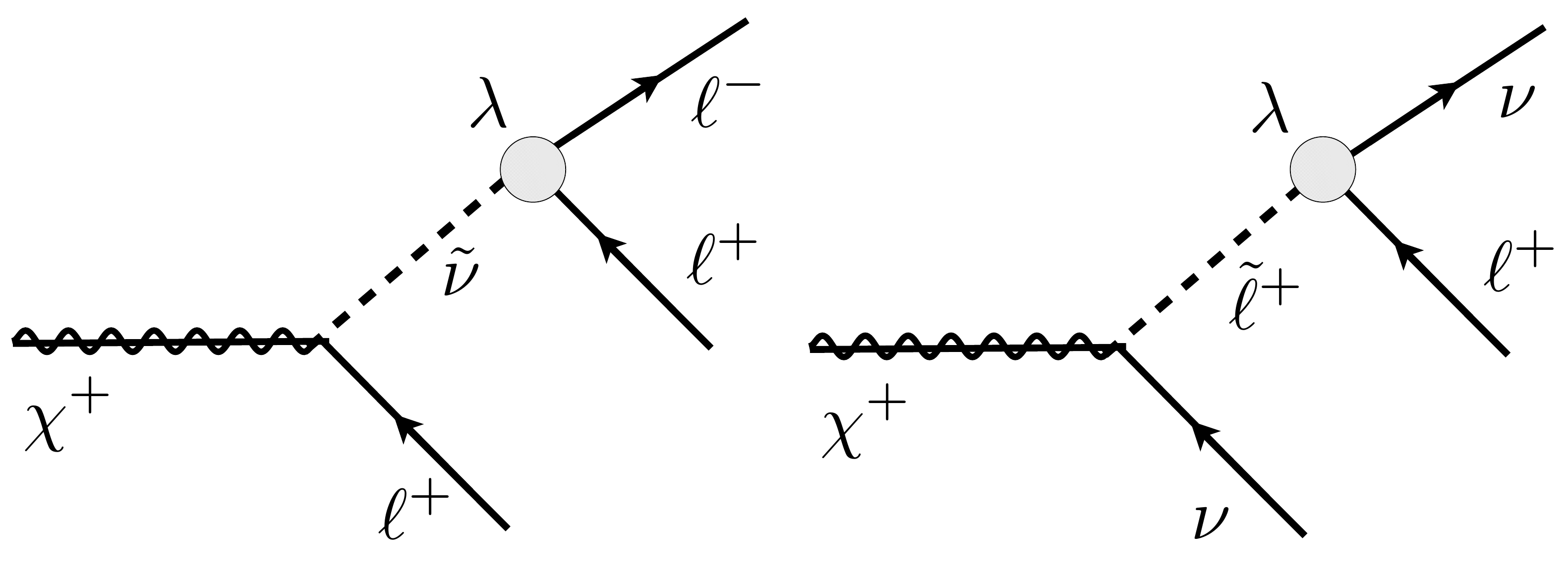}
\caption{\label{fig:RPVFeyn}  
Feynman diagrams for decays of a chargino LSP in the $R$-parity violating scenario.  The blob represents a tiny RPV coupling $\lambda$, which mediates the decay of the chargino to three charged leptons, or to one charged lepton and two neutrinos. We have suppressed lepton flavor indices, but in the text we only consider $\lambda_{323}$ to be non-zero.}
\end{figure}

A stop or stau NLSP can be meta-stable decaying to a gravitino or axino for sufficiently high scales of $\langle F \rangle$ and $f_a$ respectively, with the dominant two-body decays insensitive to the other parameters of the model. Finally, we also consider a simple $R$-parity violating (RPV) scenario with a chargino LSP.  In order to explore signals without an invisible WIMP, as well as those with a lepton rich final state, we study a model with a single dominant $\lambda LLE^c$ RPV coupling, which for wino mass $M_2 \sim 500{\rm~GeV}$ and slepton masses $m_{\tilde\ell} \sim{\rm~TeV}$ yields observable out-of-time decays when $10^{-12} \lesssim \lambda \lesssim 10^{-8}$ \cite{Shirai:2009fq}. This decay is pictured in Figure~\ref{fig:RPVFeyn}.

\subsection{Results}

As discussed in \Sec{sec:observation}, we can classify the topology of each observed event by $n_j$, the number of hard jets observed and $n_\mu$, the number of isolated hard muons observed. We require $n_j \geq 1$ to suppress calorimeter noise and cosmic ray backgrounds. The branching fractions to SM decay modes can be translated into branching fractions for the different observable decay topologies.  
For an MMCP mass of order 1 TeV and a large splitting between the MMCP and invisible particle's mass, jets from primary ($u$, $d$, $s$, $c$, $b$) quarks will be collimated and reflect the kinematic properties of the primary particle.\footnote{Muons from bottom and charm decays could be associated with jets to identify the flavor structure of of the decay, but here we will simply treat these muons as part of the jet.} Heavy particles ($W$, $Z$, $h$, $t$) will be produced with small boosts ($\gamma\sim2-5$), and although some directional information will be preserved, their decay products will lead to separated jets and muons. 

Cascade decays in the WIMP sector typically produce $Z$'s, $W$'s, and lighter SM particles. If the splittings are comparable to the WIMP mass, this can produce additional hard jets and muons in the event. Although such a cascade allows information to be determined about the WIMP sector particles, it also complicates the identification of the primary decay products. Fortunately, the cascade particles are typically softer than the primary decay particles and lead to higher multiplicity events, which can be exploited to explore the two sectors separately. 

In \Tab{tab:topologies}, the branching fractions to different decay topologies are shown for our benchmark models. Because the overall event rate is a priori unknown, we consider only the branching ratios between observable states ($n_j \geq 1$).

\begin{table}
\begin{center}

 \begin{tabular}{c|c|c||c|c|c|c||c|c}
Mode & $E_j$(GeV) & $\Delta\Theta$ & 1j & 2j & 3j & $\geq$4j & 1$\mu$ & 2$\mu$  \\
\hline
\multirow{3}{*}{A1: $\tilde g \rightarrow qq'\chi_i^{0,\pm}$}
			& 50  & $10^\circ$ & 1\% & 28\% & 26\% &  45\% & 13\% & 2\% \\
			& 50  & $30^\circ$ & 2\% & 30\% & 31\% &  37\% & 12\% & 2\%  \\
			& 50  & $60^\circ$ & 5\% & 43\% & 41\% &  11\% & 9\% & 1\%  \\
			& 20  & $30^\circ$ & 1\% & 21\% & 18\% &  60\% & 12\% & 1\%  \\
\hline
A2: $\tilde g \rightarrow qq'\chi_i^{0,\pm}$ & 50 & $30^\circ$	 & 2\% & 33\% & 31\% &	 33\% &	 13\%  &	 1\% \\
\hline
B: $\tilde g \rightarrow g\chi_i^0$ & 50 & $30^\circ$ & 35\% & 31\% & 32\% & 2\% & 3\%  & 3\%  \\
\hline
C: $\tilde t_1 \rightarrow \tilde G t$  & 50 & $30^\circ$ & 35\% & 40\% & 25\% & 0\% & 8\% & 0\% \\
\hline
D: $\tilde\tau_1 \rightarrow \tilde G \tau$ & 50 & $30^\circ$ & 100\% & 0\% & 0\% & 0\% & 0\%  & 0\% \\
\hline
E: $ \chi^+ \rightarrow \tau^-\tau^+\mu^+/\tau^+\nu_\tau\nu_\mu / \mu^+\nu_\tau\overline{\nu}_\tau $ &	50	& $30^\circ$	& 90\% &	 10\% & 0\% & 0\% &	 42\% &	 8\% \\
\end{tabular}
\caption{\label{tab:topologies} The branching fractions to final state topologies for several MMCP benchmarks. \textbf{Model A}: Split SUSY three-body gluino decay $\tilde g \rightarrow qq'\chi_i^{0,\pm}$  summed over branching to all neutralinos, charginos, and SM states (see \Fig{fig:GluinoFeyn}). A1 corresponds to a LSP with large Bino fraction, with low-energy mass parameters $\mu=200{\rm~GeV}$, $M_1=130{\rm~GeV}$, $M_2=260{\rm~GeV}$, and $m_{\tilde g} = 1{\rm~TeV}$, and A2 is a mostly higgsino LSP with the same low-energy parameters except $\mu=-120$.  \textbf{Model B:} Split SUSY two-body gluino decay $\tilde g \rightarrow g\chi_i^0$, same low energy parameters as for three-body decay A1. \textbf{Model C:} Stop NLSP two-body decay to gravitino LSP for $m_{\tilde t}=600$ GeV and $m_{\tilde G}=10$ GeV. \textbf{Model D:} Stau NLSP two-body decay to gravitino LSP for $m_{\tilde\tau}=150$ GeV and $m_{\tilde G}=10$ GeV. For (C) and (D) the small $\mathcal{O}(1\%)$ branching fractions to primary three-body decays \cite{Freitas:2011fx} are ignored. \textbf{Model E:} $R$-parity violating (RPV) decay of a $\sim400{~\rm GeV}$ higgsino-like chargino LSP through the superpotential operator $L_3L_2E^c_3$ (see \Fig{fig:RPVFeyn})  in the MSSM with weak-scale parameters $M_1 = -800{\rm~GeV}$, $M_2 = 1600{\rm~GeV}$, $\mu = 400{\rm~GeV}$ and a GMSB-like scalar sector with masses $\sim 3{~\rm TeV}$. Numerical calculations performed using MadGraph 5 \cite{Alwall:2011uj}, BRIDGE \cite{Meade:2007js}, and SUSYHIT \cite{Djouadi:2006bz}.}
\end{center}
\end{table}

We begin by estimating the order of magnitude of the systematic uncertainties in the measurement by varying the energy cuts and choice of angular resolution, as shown in the first rows of Table~\ref{tab:topologies} for the three-body decay of the gluino. Varying the angular resolution from $\Delta\theta=10^\circ$ to $\Delta\theta=30^\circ$ changes the muon and low jet multiplicity branching fractions by $\mathcal{O}(0.01)$.  The high jet multiplicity branching fractions vary by $\mathcal{O}(0.1)$ because there is a greater probability of jets overlapping for higher multiplicity modes, leading to greater uncertainty in identifying branching fractions. Varying the jet energy cut from $E_j = 50{\rm~GeV}$ to $E_j = 20{\rm~GeV}$ changes the low jet multiplicity and isolated muon branching fractions by $\mathcal{O}(0.01)$ and the high jet multiplicity branching fractions by $\mathcal{O}(0.1)$. The effects of different angular and energy cuts on the other benchmark models were similar and are not shown. Varying $\Delta\theta=30^\circ$ to $\Delta\theta=60^\circ$ has a much greater effect and thus distinguishing models with  an experimental performance of $\Delta\theta\sim 60^\circ$ would be considerably more difficult. We therefore focus on the case of $\Delta\theta=30^\circ$ and estimate that systematics uncertainties prevent signatures from being distinguished by muon and low multiplicity jet branching fractions that differ by less than $\mathcal{O}(0.01)$ and by high jet multiplicity branching fractions that differ by less than $\mathcal{O}(0.1)$. 

We can now attempt to distinguish different models. The first observation is that decay modes with different numbers of primary colored particles can easily be distinguished. For a given decay, the number of colored particles sets the minimum number of jets that will be present, regardless of the details of decays of secondary particles and WIMP sector cascades. For example, in Split SUSY, a long-lived gluino can dominantly decay either to a three-body mode or two-body mode through a loop as shown in Figure~\ref{fig:GluinoFeyn}.  Depending on the region of parameter space, either of the modes can dominate~\cite{Gambino:2005}, and these two modes correspond to (A1) and (B) in Table~\ref{tab:topologies}.  In model (A1), the branching fraction to 1j is only $2\%$, with the small fraction coming from the second primary jet being too soft or collinear with the first jet. In (B) on the other hand, there is only one primary colored particle, and so the branching fraction to $1j$ is $36\%$. Therefore only $\mathcal{O}(10)$ observed decays are necessary to statistically distinguish (A) from (B) at the $95\%$ confidence level. Model (A1) can be distinguished from model (C) in the same way, and this method is fairly insensitive to the exact energy cuts and angular resolutions $\Delta \theta$ used. 

There is also a region of the Split SUSY parameter space where the two and three-body decays (A1) and (B) have comparable branching fractions. In this case, a measurement can be made to determine the three-body branching fraction $\Gamma_{A1}$, which can be used to determine the UV scale of the theory \cite{Toharia:2006}. With  $\mathcal{O}(100)$ events observed, $\Gamma_{A1}$ can be determined to an accuracy of about $\pm 10\%$. This possibility is not restricted to Split SUSY; two and three-body decays can naturally compete in any model where three-body decays are mediated at a high scale and induce two-body decay at loop levels, and a similar measurement will be possible in such scenarios.  Of course not everything about Split SUSY can be revealed by these methods. For instance, models (A1) and (A2) differ only in the composition and spectrum of the neutralinos and charginos, (A1) having a mostly bino LSP and (A2) mostly higgsino. The primary observable of two hard jets due to the primary gluino decay is unchanged, and the changes in the cascade decays do not lead to a significant difference in branching fractions. Likewise, because the form of the primary decay vertex is identical in (A1) and (A2), these two scenario will also be degenerate in the kinematic distributions discussed in  \Sec{sec:effectivetheory} .

The only observable decay for model (D) involves the hadronic tau decay to a single jet, thus (D) can be ruled out by the observation of a significant fraction of higher jet multiplicity events. In particular it can be distinguished from (A), (B), and (C) with $\mathcal{O}(10)$ observed events. This conclusion is insensitive to the mass of the stau, as long as it is sufficiently massive that the jets can pass the triggers and cuts.

Another observation is that considering both muons and jets in the final state topologies is necessary to distinguish the widest variety of models. For example, it might be expected that the two-body decays of a stop (C) and gluino (B) would have fairly different branching fractions to higher jet multiplicity states. However, the WIMP cascades in (B) and the secondary top quark decay in (C) lead to nearly degenerate branching fractions to different jet multiplicities for the two models. Fortunately, the two scenarios have different branching fractions to final states containing muons. With $\mathcal{O}(100)$ observed decays, models (B) and (C) can be distinguished by their branching fraction to decays containing a single isolated muon. Likewise, the $R$-parity violating decay through the lepton-number violating operator (E) is easily distinguished by its large branching fraction to events containing hard muons.

Based on these analyses, we conclude that observing the branching fractions to different jet and muon multiplicities of $\mathcal{O}(10-100)$ late decay events is sufficient to distinguish many MMCP scenarios, in particular those which differ in the color representation of the meta-stable particle or the number of leptons produced in the decay. This conclusion relies on the assumptions discussed earlier that an angular resolution of $\Delta\Theta\sim30^\circ$ can be obtained, and that background rate from cosmics can be reduced to $\mathcal{O}(1)$ events. Comparing to \Tab{tab:eventrate}, this corresponds for instance to direct pair production cross sections of $40-400{\rm~fb}$ and thus a mass reach of roughly $1.0-1.3$ TeV for a color octet fermion.

For some measurements and models, it can be competitive to identify the MMCP and its charges directly in production events.  Looking for very high momentum tracks in the muon chamber could be one of the first signs of a discovery of a CHAMP~\cite{Gupta:2007ui}. Mass measurements can be performed with great accuracy from time-of-flight measurements, with for example better than $1\%$ accuracy for a gluino up to 1.5 TeV \cite{Kilian:2004uj}. If another new particle decays promptly to the MMCP, the couplings and mass of the MMCP can sometimes be revealed \cite{Choudhury:2008gb, Biswas:2009zp, Biswas:2009rba, Feng:2009yq,Ito:2009xy, Biswas:2010cd,Ito:2010un}. For instance, for certain motivated spectra a fraction of stop squark MMCP production will come from gluino decays, and the identity of the stop MMCP can be determined with similar reach to our proposed methods from the presence of top quarks in these events \cite{Choudhury:2008gb}. More generically, standard model particles radiated from the MMCP in direct production events probe the MMCP couplings \cite{Chang:2011jk, Luty:2011hi}. For example, $\mathcal{O}(1)$ yukawa couplings of a colored MMCP to the Higgs can be constrained with $\sim20,000$ direct production events \cite{Luty:2011hi}; this could for example distinguish a 1 TeV gluino from a 4th generation vector-like quark, giving this search comparable reach to our proposal for identifying gluinos. Another proposal studied in less detail in the literature is to exploit the differences in energy losses to distinguish color triplet MMCPs from color octets \cite{Buckley:2010fj}. 

These and similar proposals for measuring the properties of other MMCPs in production events are complementary to our proposed measurements in several ways. For colored MMCPs, while we are sensitive to the interaction of R-hadrons in the detector in estimating the mass reach of our methods, no knowledge of these interactions is required to perform the actual measurement. In contrast, measurements of production events are sensitive to the in-flight detection and tracking of R-hadrons, and therefore to uncertainties in modeling the interactions and spectra of R-hadrons. In particular, most production searches rely on detecting the MMCP in the muon system, and their efficiency will be greatly reduced if charge suppressing inelastic interactions take place as the R-hadron propagates through the calorimeters \cite{Mackeprang:2010}. More generally, our measurements are not sensitive to the details of the production event, including any other new particles contributing to the MMCP production. Finally, for MMCPs that do not decay in-flight, the nature of the decay, including the Lorentz structure of the decay and any produced new particles which couple very weakly to the standard model, can only be probed by observing the details of the out-of-time decays.

\section{Determining Lorentz Structure}
\label{sec:effectivetheory}

As we saw in \Sec{sec:topologies}, many MMCP scenarios can be distinguished by the topology of the decay. However, some scenarios could still be difficult to distinguish, for instance color octet SIMPs of different spins could have similar standard model decay modes. If an MMCP is observed to have dominantly three-body decays, then observing the kinematic distribution of the primary decay particles provides a further test of UV and TeV physics by helping determine the Lorentz structure of the couplings and the spins of the MMCP and WIMP.\footnote{In the case of two-body decays, the kinematics are trivial and contain no information.}  In this section, we study how this information can be obtained from the kinematic properties of the final state muons and jets in an MMCP decay.

\subsection{Decay Operators}

Most of the models discussed in Section~\ref{sec:topologies}  yield metastable particles $X$  because the stability of $X$ is violated by physics at some high scale $\Lambda$. We can therefore integrate out the high scale physics and study the MMCP decay in each scenario through the resulting effective operator. In particular, we would like to answer the question: assuming the decay topology is compatible with a given MMCP scenario, can measurements of the late decay kinematics at the LHC provide a meaningful test of or rule out other possible decay operators? To answer this question, our strategy is to compare the kinematic distributions for a given scenario to a wide variety of other possible decay operators yielding similar decay topologies but different kinematic distributions.

The possible Lorentz structures and observable kinematics of the operators depend only on the spins, so we have listed the operators by the MMCP and WIMP spins. \Tab{tab:FFoperators} lists operators involving two SM fermions $(f=Q_L, u_R, d_R, \ell_L, e_R)$. Three-body decays to final states involving SM bosons are also possible, but have not been included because they typically can be distinguished from our reference scenarios using the methods of \Sec{sec:topologies}, and considering them would not significantly change our conclusions.  In listing these operators, we adopt the following notation. The letter X will of course denote the metastable particle, with $X$ a fermion, $\tilde{X}$ a scalar and $\mathcal{X}$ a vector. Since many motivated cases involve a DM candidate, we will assume such an invisible particle Y is present in the decay, but our results are easily generalizable to the case where all primary particles produced are visible, e.g. an RPV decay.   Of course, this list is not meant to be exhaustive -- the operators have however been chosen to give a wide representation of the class of dimension five and six operators which can emerge naturally from heavy physics. In particular the operator $O_{S2}^{ff}$ corresponds to the angular distribution between the quarks in the three-body decay of the long lived gluino in Split SUSY (see \Sec{sec:models}). The same operator also corresponds to the angular distribution between the tau-jets in the three-body RPV chargino decay in the limit $m_{\tilde \ell} \gg m_{\chi^+_1}$.\footnote{The kinematics of the muon from the RPV chargino decay could certainly be used to improve the measurement of the tau-jet angular distribution, but for our purposes here it is sufficient to simply treat it as the invisible particle $Y$.}

\begin{table}
\begin{center}
 \begin{tabular}{c||c|c}
	$J_{\rm MMCP} \times J_{\rm WIMP}$ & \multicolumn{2}{l}{Decay operators ($ff$ modes)} \\
\hline\hline
\multirow{2}{*}{$0\times 0$}
					& $O_{S}^{ss}$		&	$\Lambda^{-1}( \bar f_R^2  f_L^1 )  (\XS \, \YS)$ \\
					& $O_{V}^{ss}$		&	$\Lambda^{-2}( \bar f_L^2 \gamma^\mu  f_L^1 )  (\XS \partial_\mu \YS - \YS \partial_\mu \XS)$ \\
\hline
  \multirow{5}{*}{$\frac{1}{2}\times\frac{1}{2}$} 	
					&  $O_{S1}^{ff}$ 		& $\Lambda^{-2}( \bar{f}^2_R f^1_L ) (\bar  Y X )$ \\
					&  $O_{S2}^{ff}$ 		& $\Lambda^{-2}( \bar {f}^2_R X ) (\bar Y f^1_L )$ \\
					&  $O_{V1}^{ff}$		& $\Lambda^{-2}(\bar {f}^2_L \gamma^\mu f^1_L ) (\bar Y \gamma_\mu X )$ \\
					&  $O_{T1}^{ff}$		& $\Lambda^{-2}( \bar {f}^2_L \sigma^{\mu\nu} f^1_R ) (X \sigma_{\mu\nu} Y )$ \\
  					&  $O_{T2}^{ff}$		& $\Lambda^{-2}( \bar {f}^2_L \sigma^{\mu\nu} X ) (\bar f^1_R \sigma_{\mu\nu} Y )$\\
\hline
\multirow{5}{*}{$1\times1$} 
					& $O_{S}^{vv}$		&	$\Lambda^{-1}( \bar {f}^2_R f^1_L )  (\XV_\mu \, \YV^\mu)$ \\
					& $O_{T}^{vv}$		&	$\Lambda^{-1}( \bar{f}^2_R\sigma^{\mu\nu}f^1_L  )  (\XV_\mu \, \YV_\nu)$  \\
					& $O_{V1}^{vv}$		&	$\Lambda^{-2}( \bar f^2_L \gamma^\mu f^1_L ) ( \YV_\nu  \partial^{\nu} \XV_\mu )$ \\
					& $O_{V2}^{vv}$		&	$\Lambda^{-2}( \bar f^2_L \gamma^\mu f^1_L ) ( \XV_\nu  \partial^{\nu} \YV_\mu )$ \\
					& $O_{V3}^{vv}$		&	$\Lambda^{-2}( \bar f^2_L \gamma^\mu f^1_L ) ( \XV_\nu  \partial_{\mu} \YV^\nu - \YV_\nu  \partial_{\mu} \XV^\nu)$ \\
\hline
\multirow{3}{*}{$1\times 0$}
					& $O_{S}^{vs}$		& 	$\Lambda^{-2}( \bar{f}^2_R f^1_L ) ({\XV}^\mu \,\partial_{\mu} \YS)$ \\ 
					& $O_{V}^{vs}$		&	$\Lambda^{-1}( \bar f^2_L \gamma^\mu f^1_L )  (\XV_\mu \, \YS)$ \\
					& $O_{T}^{vs}$		&	$\Lambda^{-2}(\bar {f^2_R}\sigma^{\mu\nu}f^1_L ) (\XV_\mu \,\partial_{\nu} \YS - \XV_\nu \,\partial_{\mu}\YS)$ \\
\end{tabular}
\caption{\label{tab:FFoperators} Three-body decay operators with a WIMP and two SM fermions in the final state, listed by MMCP and WIMP spin.  Because of the chiral nature of the SM gauge group, only chiral couplings of the SM fermions are considered. The operators for $0\times 1$ are the same as $1\times 0$ with the WIMP and MMCP interchanged and are denoted $O_{(~)}^{sv}$. For all operators the addition of the Hermitian conjugate is implied.}
\end{center}
\end{table}

\subsection{Angular Distributions}

As described in \Sec{sec:observation}, we take the angle $\theta$ of each jet to be measured within a Gaussian distributed error of $\Delta\theta$. We wish to relate this measurement to the distribution $dN/d\theta$ of the opening angle between the two primary SM particles in a decay. Because the MMCP decays at rest, the initial center of mass frame is known and therefore this distribution will directly carry information about the decay operator. The ideal situation for making this measurement would be a decay to lighter quarks ($u$, $d$, $c$, $s$, $b$) or $\mu$, the observable signatures of which will reflect directly the primary particle kinematics up to the uncertainties due to detector resolution as discussed in \Sec{sec:observation}.

\begin{figure}
\center{
\includegraphics[width=19cm]{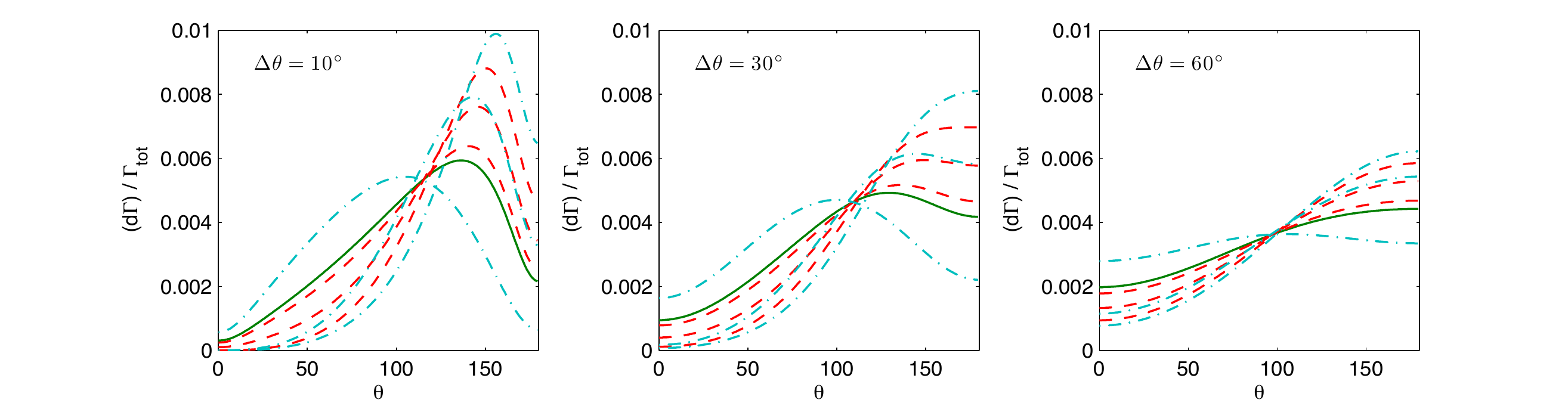}}
\caption{\label{fig:X8split_rel_smear} Selected angular distributions $dN/d\theta$ for angular resolutions of $\Delta\theta = 10^\circ,~30^\circ,~{\rm and}~60^\circ$ from left to right for highly relativistic WIMP ($m_{X}/m_{Y}=10$). Solid (green): Reference Split SUSY three-body distribution $O_{S2}^{ff}$. Dashed (red): other representative allowed operators with same MMCP and WIMP spin and gauge representation ($O_{T1}^{ff}$, $O_{T2}^{ff}$, $O_{S1}^{ff}$ from top to bottom on LHS).  Dot-dashed (blue): operators allowed for same gauge representation but different spins ($O_{T}^{vv}$, $O_{S}^{vv}$, $O_{S}^{ss}$ from top to bottom on LHS). All angular distributions shown were generated using the CompHEP and LanHEP software packages \cite{CompHEP,LanHEP}.}
\end{figure}

The kinematic distributions for some of the representative operators in Table~\ref{tab:FFoperators} are plotted in Figure~\ref{fig:X8split_rel_smear} for the different benchmark values of the angular resolution $\Delta\theta$. The normalized angular distributions are plotted, since the overall event rate is a priori unknown. We do not apply any energy cuts to these distributions, but as we will see in a realistic scenario this does not significantly affect the distribution. As is evident from these plots, there are order one differences in the distributions and hence it should be possible to discriminate between operators by using simple counting statistics such as the fraction of decays occurring between two angular intervals. It is clear that we have discriminating power even with the rather coarse angular resolution assumed for the LHC detectors in Section~\ref{sec:observation}. When $\frac{M_X - M_Y}{M_Y}\gg 1$, the distribution will be insensitive to the exact values of the WIMP and MMCP masses. Otherwise, we assume that the MMCP mass can be measured in production events and that the splitting can be determined well enough from the statistical distribution of total energy deposits to allow the appropriate distributions to be compared. 

\begin{figure}
\centering
\includegraphics[width=12cm]{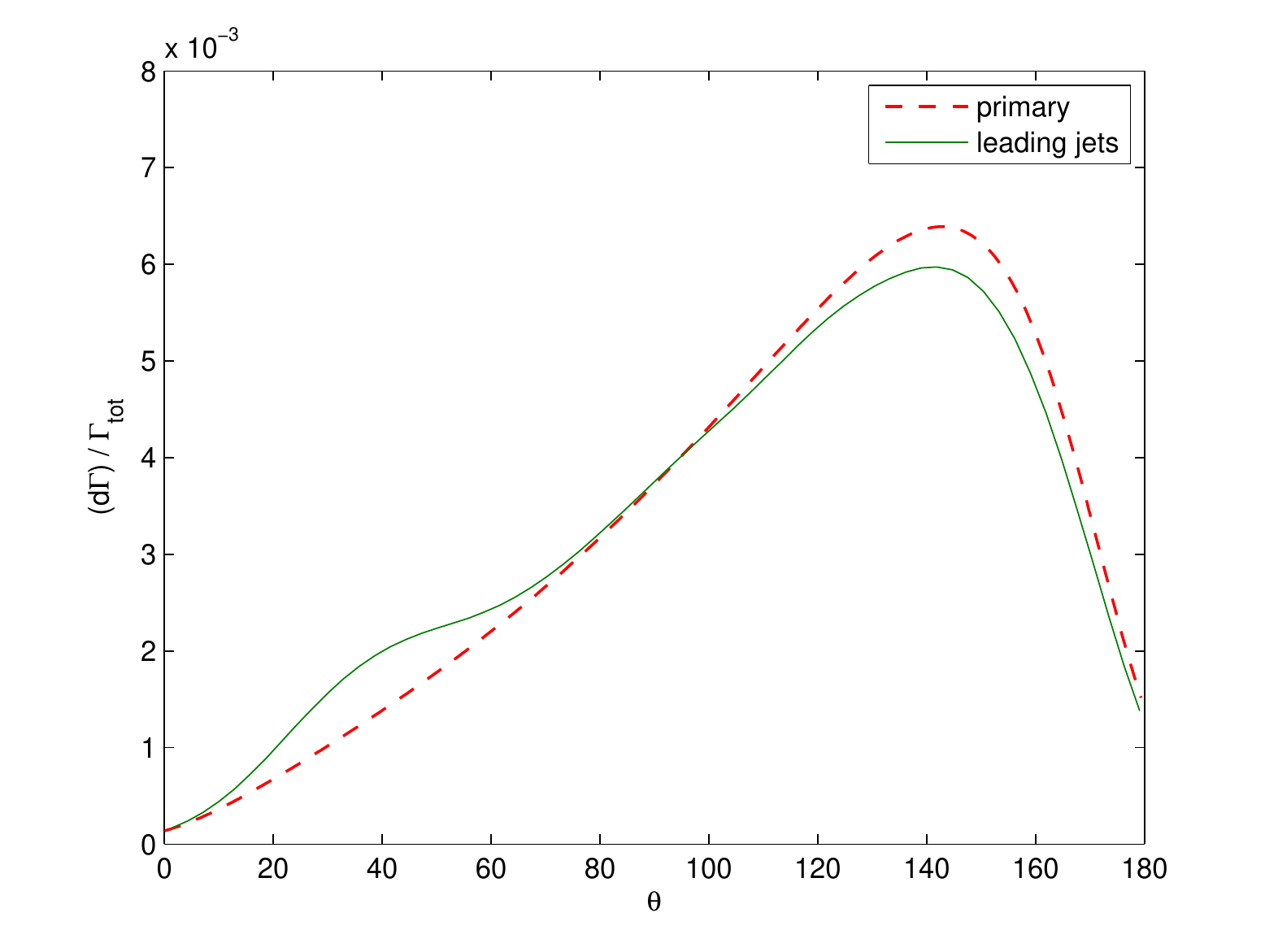}
\caption{\label{fig:top-jets} Angular distribution for the two leading energy jets produced in gluino decay of Model A from Table~\ref{tab:topologies} including all decay modes and secondary decays (solid) compared to the angular distribution of primary quarks in the channel where only two light quarks and the WIMP are produced (dashed). The secondary decays of top quarks and chargino/neutralinos are primarily responsible for the differences in the two distributions.}
\end{figure}

In scenarios where the event has a higher multiplicity of jets or muons due to heavy decaying primary particles ($W$, $Z$, $h$, $t$) or cascade decays in the WIMP sector, it is more difficult to reconstruct the primary particle angular distribution. One possible strategy is to group jets to try to reconstruct the initial primary particle; for instance jets from $W$ decay will still have an angular correlation despite the low boost factor. Another strategy, if limited jet energy resolution is possible, would be to determine the angular distribution between the two highest energy jets in each event. Unless the secondary decays have splittings comparable to the primary decay, this distribution will tend to reflect the kinematic distribution of the primary quarks produced in the decay. To illustrate this, \Fig{fig:top-jets} compares the normalized angular distribution between the idealized case when only primary light quarks and a WIMP are produced in the gluino decay model A1 (corresponding to $O_{S2}^{ff}$) with no energy cuts, to the true distribution  of the two leading energy jets after the heavy primary particle decays and WIMP cascade decays with an energy cut $E_j > 50$  GeV. As can be seen, the distribution still carries much of the original kinematic information. Therefore for the remainder of this analysis we will simply consider the distributions of the two primary SM particles produced in an event.

\subsection{Results}

A simple way to quantify the difference between the angular distributions of two operators is to ask how many three-body decays would need to be observed to distinguish one from another. The average number of decays necessary to distinguish two distributions can be estimated using the Kullback-Leibler distance (see \cite{Csaki:2007} for a description in a similar context),
\begin{equation}
\label{eq:KLdist}
N(O_1, O_2) = \frac{\log R}{KL(O_1, O_2)}; \quad\quad\quad KL(O_1,O_2) = \int d\theta \log\left(\frac{dN_1/d\theta}{dN_2/d\theta}\right)\frac{dN_1}{d\theta},
\end{equation}
where $dN/d\theta$ are the normalized angular distributions, and $R$ is the required confidence that the distribution does not correspond to the operator $O_2$, given that the true distribution is from $O_1$.

\begin{figure}
\centering
\includegraphics[width=12cm]{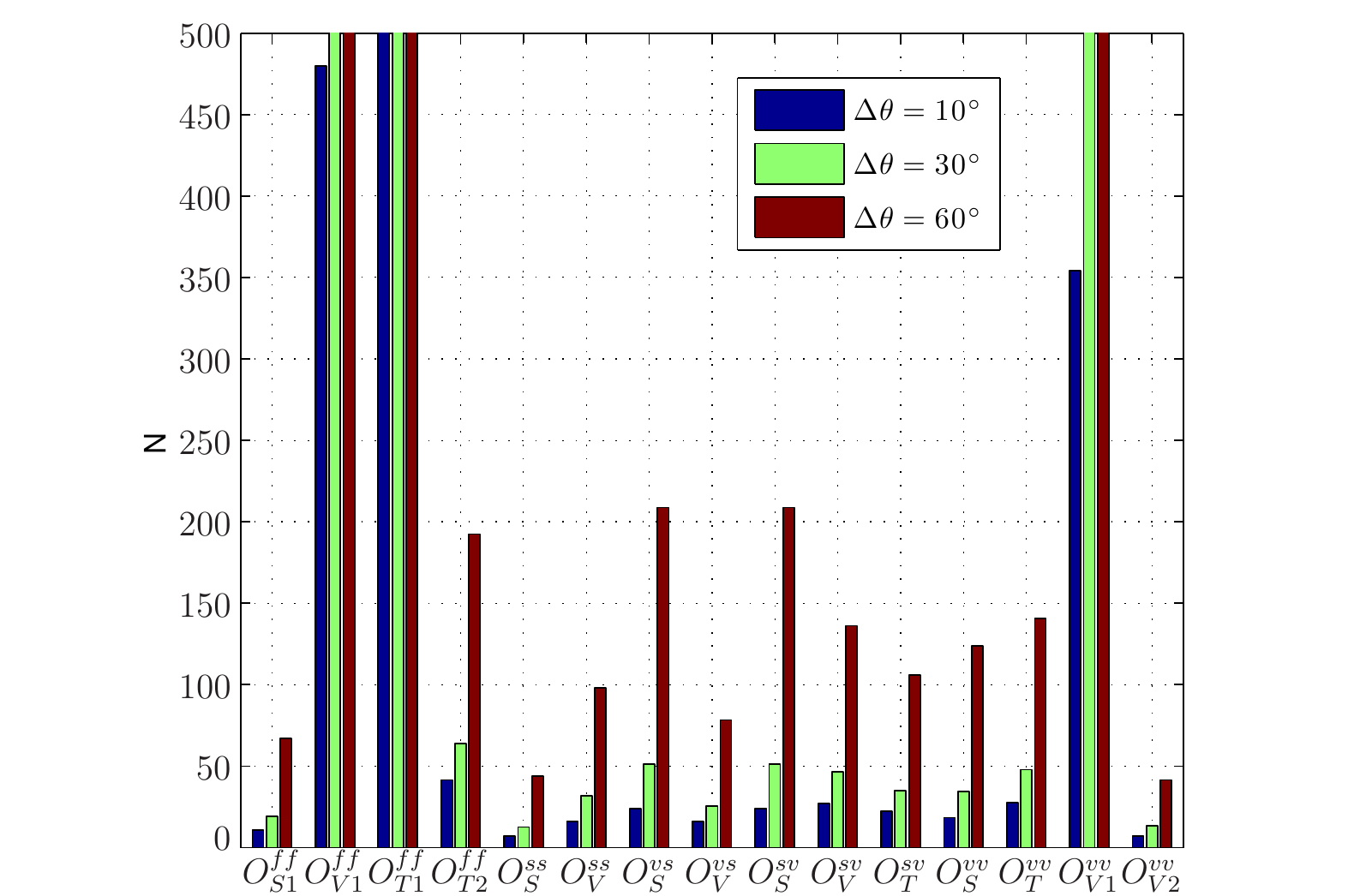}
\caption{\label{fig:FFevents} Average number of observed decays necessary to distinguish decay operators at 95\% confidence level from the reference operator $O_{S2}^{ff}$, from \Eq{eq:KLdist} using the distributions for $m_{X0}/m_{X8}=10$. For each operator, from left to right the angular resolution is $\Delta\theta=10^\circ,~30^\circ,~60^\circ$.}
\end{figure}

Using this statistic, in Figure~\ref{fig:FFevents} we estimate the number of events needed to differentiate the operators in Tables~\ref{tab:FFoperators} from the reference Split SUSY gluino / RPV chargino distribution coming from $O_{S2}^{ff}$. Note that in most cases, we only need $\sim \mathcal{O}\left(100\right)$ observed three-body decays to distinguish various operators even when the angular resolution  $\Delta \theta_{12}\approx 60^\circ$. With better angular resolution  $\Delta \theta_{12}\approx 10^\circ$, we can distinguish most operators with  $\sim \mathcal{O}\left(10\right)$ stopped particles. In \Tab{tab:testing} these numbers are converted to the necessary production cross section and corresponding mass thresholds at the 14 TeV LHC for the applicability of these tests to various UV MMCP scenarios, using the factors from \Sec{subsec:signal}.  Assuming only direct production of MMCPs, with the most optimistic $\Delta\Theta$ resolution the 14 TeV LHC has a mass reach of $1.2$ TeV ($400$ GeV) for testing the identity of a gluino (chargino) MMCP through the angular distribution of decays.
\begin{table}
\begin{center}
 \begin{tabular}{c||c|c|c}
				 &  $N$  & $\sigma$ & $M$ \\
	\hline
	$\Delta\theta_{12} = 10^\circ$ & 23 & 90 fb (60 fb)& $\sim1.2$ TeV  ($\sim400$ GeV) \\
	$\Delta\theta_{12} = 30^\circ$ & 41 & 160 fb (100 fb) & $\sim1.1$ TeV ($\sim350$ GeV) \\
	$\Delta\theta_{12} = 60^\circ$ & 130 & 520 fb (370 fb) & $\sim0.9$ TeV ($\sim250$ GeV)\\
\end{tabular}
\caption{\label{tab:testing}  $N$ is the median value of the set of all $N_i$, the number of three-body decay events necessary to distinguish the reference operator $O_{S2}^{ff}$ from an operator $O_i$ in Table~\ref{tab:FFoperators}. Also shown are the approximate necessary production cross section and direct production mass reach for this number of events for a color octet SIMP (SU(2) doublet CHAMP) respectively as discussed in \Sec{sec:topologies}.}
\end{center}
\end{table}

The spin of the MMCP could also be determined from production events. To compare to our methods, we note that $\mathcal{O}(100)$ observed out-of-time decays in the ECAL, corresponding to $\mathcal{O}(40,000)$ total pair production events, can distinguish a variety of different decay operators. For colored MMCPs, Ref. \cite{Buckley:2010fj} found that a similar magnitude of direct production events would need to be observed to distinguish a scalar, fermion, and vector MMCP from one another by observing the angular distribution of the production. For color singlet MMCPs, Ref. \cite{Rajaraman:2007ae} found that the angular distribution in pair production events for a $\mathcal{O}(100)$ GeV scalar can be distinguished from a fermion with only $\mathcal{O}(3000)$ total production events. A major advantage of direct production events over our methods is that the spin can be measured directly, rather than constrained through the observed decay operator. Furthermore, in the case that the decay is dominantly two-body, production events may be the only way to access the spin information. On the other hand, measuring spin with direct production angular distribution measurements becomes more difficult if other new particles contribute significantly to the production cross section.  Furthermore, the angular distribution in decay events can provide information beyond the spin of the MMCP, for example it can distinguish different decay mechanisms involving particles of the same spin.

\section{Conclusions}

New metastable particles occur in many extensions of the Standard Model. The gravitino or axino in the MSSM, the gluino in Split SUSY, or small $R$-parity violation give well-motivated examples of such decays.  Frequently the metastable particle is charged or colored so, if light enough to be produced at the LHC, some fraction will stop in the detectors.  These MMCPs then decay out of time, giving an observable and striking signature.  Searches for such events are currently underway.  In this work, we considered how measurements of these late decays in the LHC detectors could go beyond a detection to explore the properties of both the MMCP and perhaps even a dark matter particle produced in the decay. With reasonable assumptions about the LHC performance, the observed couplings to standard model particles can give insight into the gauge properties of the new particles, and, when a dominant decay mode is three-body, the spin properties can also be constrained through observed angular distributions of decay products.

Additionally, the decay of such a long-lived particle is often due to an accidental symmetry being broken in the UV.  In this case the structure of the decay operator is determined by the UV physics giving rise to the decay.  Our suggested measurements would then provide significant hints of the UV physics, far above the scales the LHC can probe directly, by determining the dominant couplings to standard model particles, and in the case of three-body decays differentiating operators of different Lorentz structure.  For example, although the squarks in Split SUSY are far above the TeV scale, they could be indirectly ``observed" in this way through the out of time decays of the gluino.

We found that $\mathcal{O}(10-100)$ observed late decays originating in the electronic barrel calorimeter of CMS or ATLAS is sufficient for discriminating many different MMCP candidates, see Figure~\ref{fig:FFevents} and Table~\ref{tab:testing}.  Taking into account only direct production as in Table~\ref{tab:eventrate}, this corresponds for instance to a mass reach of 1.0-1.3 TeV for a gluino at the 14 TeV LHC, 400-600 GeV for a stop squark, and 300-500 GeV for a nearly-degenerate chargino. In particular, the measured decay topologies can distinguish a variety of motivated models, especially those differing in the color representation of the MMCP. For the case of Split SUSY, this result was found to be insensitive to the composition of the neutralino. We also found that the relative branching fractions of decay modes differing in final state colored particle multiplicity can be measured to $\sim 10\%$ accuracy using the example of the two and three-body decay modes of the gluino in Split SUSY. Furthermore, we showed that even in cases with degenerate decay topologies, if there is a significant branching fraction to three-body decay modes, then the kinematics of the decay can provide a nontrivial test of the MMCP and WIMP spins and the Lorentz structure of their couplings.

Discovery searches place lower limits for example at $\sim900-1100$ GeV for a color octet fermion (gluino) and $\sim600-700$ GeV for a color triplet scalar (stop) depending on the R-hadron model~\cite{Chatrchyan:2012}.\footnote{These new limits were released after this work first appeared.} Our proposed measurements are therefore relevant to any colored MMCP discovered in the remainder of the 7 or 8 TeV LHC run, or the beginning of the 14 TeV run, although for a scalar triplet MMCP further optimization or more luminosity than our benchmark may be required. Our proposed measurements are also relevant to non-colored MMCPs, although staus with sufficient direct production cross section to be measured by our techniques are excluded by Ref.~\cite{Chatrchyan:2012}.

The measurements we have discussed can be made independently of any measurement of the MMCP properties or cross section in production events. They can therefore provide a source of information independent from other proposed measurements of the MMCP gauge and spin representation. Moreover, because the scale of physics mediating the MMCP decays is generally far above the TeV scale, late decay measurements probe physics completely inaccessible in direct production events. To make the most of the capabilities of the LHC detectors in this out-of-time decay window would require a dedicated experimental effort to understand the angular measurement uncertainties for jets on a scale of $\Delta\theta\sim30^\circ$ and to control cosmic ray backgrounds to $\mathcal{O}(1/{\rm yr})$ in the ECAL. We hope that this work has demonstrated the plausibility and utility of this effort, and that it can serve as a useful resource providing motivated theoretical benchmarks and a point of reference for comparison to more exotic proposals for dedicated detectors and upgrades targeted at such metastable particles.

\section*{Acknowledgments}

We would like to thank Cliff Cheung, Sarah Eno, Tom LeCompte, Ken Rossato, Neal Weiner, David E. Kaplan, and Andy Haas for helpful discussions. DS would like to thank the Berkeley Center for Theoretical Physics for their hospitality.  DS is supported in part by the NSF under grant PHY-0910467 and gratefully acknowledges support from the Maryland Center for Fundamental Physics. KH is supported by the NSF Graduate Research Fellowship under Grant No. DGE-0645962.  This work was supported in part by ERC grant BSMOXFORD no. 228169.

\appendix

\section{Measuring out-of-time decays in the ATLAS/CMS barrel ECAL}
\label{app:measurement}

In this section we describe and motivate a strategy for identifying out-of-time decays originating in the barrel ECAL at ATLAS or CMS and measuring the direction and multiplicity of muons and jets in the decay on an event-by-event basis.

When interactions occur at the collision point, the fine segmentation of the barrel calorimeters in the transverse $\eta\times\phi$ plane allows good reconstruction for the direction of isolated energy deposits, even without tracking information. However, the vast majority of stopped particles in ATLAS and CMS will come to rest in the central electronic and hadronic calorimeters \cite{Arvanitaki:2005nq}. 
With the decay vertex in the calorimeter, the geometry of the detector is no longer projective in the decay angles. Furthermore, the energy readout in the calorimeters is sensitive to the location and orientation of an event relative to the active and inactive components of the calorimeter cells, and it is unlikely that the visible energy can be measured reliably on an event-by-event basis\footnote{It seems likely, however, that for a large number of decays, the statistical distribution of energies could be used to roughly determine the MMCP-WIMP splitting from the average visible decay energies. } for decay vertices inside the calorimeters.

A promising observable however is the geometric pattern of energy deposition in the calorimeter cells, which could reveal the direction of the primary particles and the location of the decay vertex. Electromagnetic energy will penetrate relatively few calorimeter cells, and it is therefore unlikely that useful directional information can be obtained from pure electromagnetic energy deposits except in the small fraction of decays producing products that pass back through the inner detector or into the muon chambers. Hadronic products will however deposit their energy over several nuclear interaction lengths, which corresponds to several cells in the HCAL and many cells in the ECAL. It therefore seems likely that directional information about jets can be obtained. Another promising observable is hard muons, which will generally escape the calorimeters to pass through the muon chambers and are therefore straightforward to measure. 

Motivated by these considerations, we want to consider how jets and muons produced in late decays can be measured at the LHC detectors. 

\subsection*{Measuring Jets}

We first consider decays occurring inside the HCAL.  The radial segmentation of the calorimeters is poor at both ATLAS and CMS.\footnote{At ATLAS the electromagnetic and hadronic calorimeters are each read out in two narrow and one thick central radial shells, while at CMS the ECAL has a single radial readout, and the HCAL has two.} Because the HCAL is designed to contain radiation, all the information about radial momentum will be lost.  One could reconstruct the decay in the plane transverse to the radial direction, but the orientation of the decay relative to that plane is unknown.  This ambiguity will distort the angular distributions discussed here, making measurements much more difficult. 

We instead focus on decay vertices located in the ECAL, which have the advantage that much of the hadronic radiation will escape this calorimeter and deposit energy in other components of the detector (at ATLAS and CMS the ECAL has an annular radius of about one nuclear interaction length, while typical hadronic components deposit significant energy over a distance of about five interaction lengths~\cite{Adragna:2010zz}). These decays can be identified by a large electromagnetic fraction for the calorimeter energy deposit. Depending on the angle of the final state visible particle relative to the detector geometry, different trajectories are possible.  If the momentum is pointing radially outward, the radiation will leave the ECAL and directly enter the HCAL and further shower.  If the momentum is pointing inward, it will leave the ECAL into the central tracker, deposit some tracks, then re-enter the calorimeters in a different region. A small fraction of the events will have a very small radial momentum and stay confined to the ECAL, but here we focus on the majority of events where some radiation escapes the ECAL. The geometry is shown in Figure~\ref{fig:strategy} for a decay originating in the ECAL with two jets in the final state.  We note that this figure is schematic and the angular resolution of the calorimeters is much finer than the depiction in the figure.  The relevant geometric properties of the ATLAS and CMS detectors are summarized in \Tab{tab:ATLAS_CMS}. Also relevant is the nuclear interaction length $\lambda_I\sim 20{\rm~cm}$ for the absorbing components of the calorimeters.

\begin{table}
\begin{center}
 \begin{tabular}{p{5cm}||c|c}
						&	ATLAS	&	CMS 	\\ 
\hline
\hline
ECAL inner radius	($R_{\rm ECAL}$)	&	115 cm	&	129 cm 	\\ 
\hline
ECAL instrumented annular radius ($\Delta R_{\rm ECAL}$) & 47 cm	&	23 cm 	\\ 
\hline
HCAL inner radius	($R_{\rm HCAL}$)	&	228 cm	&	175 cm	\\
\hline
HCAL instrumented annular radius & 164 cm  & 96 cm	\\
\hline
ECAL $\eta\times\phi$ resolution	& $0.025\times0.025$		&	$0.0174\times0.0174$	\\
\hline
HCAL $\eta\times\phi$ resolution	& $0.1\times0.1$			&	$0.087\times0.087$	\\
\hline
\end{tabular}
\caption{\label{tab:ATLAS_CMS} Geometric properties of the barrel components of the CMS and ATLAS calorimeters from \cite{Bayatian:2006zz, :1999fq, Ille:1999pz, Abdullin:2008zzb, Adolphi:2008zzk, Abdallah:2008zz}. }
\end{center}
\end{table}

Due to the poor radial segmentation of the ATLAS and CMS calorimeters, the uncertainty in the measurement of $\theta_{12}$ is dominated by the radial uncertainty not the resolution in $\eta\times\phi$ plane. For particles that exit the ECAL into the central tracker, like the jet going down the page in Figure~\ref{fig:strategy}, it is simple to estimate the angular resolution achievable. The first point in the trajectory is the decay vertex, which can be located in the $\eta\times\phi$ plane through the shape of the deposit of EM energy in the ECAL. The radial uncertainty is given by the annular radius of the ECAL, $\Delta R_{\rm ECAL}$. The second point on the trajectory is the location where the energy deposit re-enters the ECAL. The characteristic angular resolution for this method is simply estimated in terms of the ECAL inner and annular radii,
\begin{equation}
 \Delta\theta \sim \frac{\Delta R_{\rm ECAL}}{R_{\rm ECAL}}.
\end{equation}
For ATLAS, $\Delta\theta\sim 25^\circ$, and for CMS it is even better, $\Delta\theta\sim 10^\circ$.

For particles like the jet going up the page in Figure~\ref{fig:strategy} which go directly from the ECAL to the HCAL, the second point in the trajectory is the center of the HCAL energy deposit, which will occur in one of the nearby cells. The uncertainty in angle can be estimated from the distance $R_{E-H}$ between the radial centers of the ECAL and and HCAL cells to be $\Delta\theta\approx \frac{\Delta R_{\rm ECAL}}{R_{\rm E-H}}$, with $\Delta\theta\sim 35^\circ$ for ATLAS and $\Delta\theta\sim 25^\circ$ for CMS.

These simplified cases and geometric estimates suggest that a resolution of $\Delta\theta\approx30^\circ$ is plausible to expect for the measurement of jet direction, and that the geometry of CMS makes a slightly better measurement possible. Of course without a detector simulation there remains uncertainty in the true capabilities of the detectors, motivating us to to also consider optimistic and conservative benchmarks of respectively $\Delta\theta\approx10^\circ$ and $\Delta\theta\approx60^\circ$.

Another source of uncertainty is the fact that colored particles undergo parton showers and hadronization and deposit energy in the form of jets.  From jet shape studies \cite{Martinez:2009aw}, on average $\sim90\%$ of jet energy will be deposited within an angular cone of opening angle $\theta_{\rm jet}\sim30^\circ$ at $E=100$~GeV.  For a 1 TeV $X$ particle, the jet energies will be somewhat higher making their opening angles moderately smaller, but this sets the rough limit on the relative angle at which two jets can be distinguished from one another, and is incidentally roughly the same as the overall angular resolution.

A more accurate method of making angular measurements might be possible through further study of the shapes of energy distributions in the calorimeter cells, as well as the possible incorporation of inner detector measurements of charged particles. In our work, other possible strategies considered had roughly the same or worse accuracy and efficiency. 

Finally, we comment on the more remote possibility of obtaining more detailed radial information from the calorimeters.  For instance, the HCAL at CMS has 17 radial segments which are an average of about 6 cm thick. The current setup of CMS integrates this into two different readouts, so only very coarse transverse information is kept.  If a long lived charged particle is discovered, however, the experiment may achieve significant gains in reconstruction of the decay points and jet direction within the HCAL by separately reading out each of the layers. 

\subsection*{Measuring Muons}
\label{subsec:muons}

Isolated muons can be produced as primary particles in an $X$ decay.  Hard isolated and non-isolated muons can also be produced in secondary decays. For a decay originating in the ECAL, the muon direction can be measured from the location in $\eta\times\phi$ of its entry into the barrel muon system, or in $R\times\phi$ plane for the endcap muon systems. The expected angular resolution is superior to to the jet angular resolution, which we take to be the limiting factor in our analysis.

\section{Stopping MMCPs}
\label{app:stopping}

When an MMCP X is produced, it is slowed down due to electromagnetic interactions with the electrons in the detector material. For a given amount of material, the energy loss due to these interactions will on average stop any X moving slower than a critical velocity $\beta$. We compute the fraction of all X produced which stop in the central portion of the ATLAS and CMS ECAL by simulating the production of different MMCPs in MadGraph 5 and applying a simple model of the detector geometry. 

The kinematic distribution and therefore the stopping fraction for a given MMCP depend on its production mechanism and mass. If $X$ is colored and produced in the primary hard process, it will often be produced near threshold (due to PDF suppression) and will be slow. This is likely in the case of heavy colored $X$s such as the gluino. For direct pair production, scalars such as the stau and stop will tend to be produced with a harder spectrum than fermions like gluinos and charginos. $X$ could also be produced as a result of cascade decays of heavier particles, in which case those particles will be more relativistic. This should be expected in the case of the long lived stau or chargino in supersymmetry, where it will be produced dominantly as a result of cascade decays from heavier colored particles. When $X$ carries color, there is an additional source of uncertainty coming from the spectrum of charged R-hadron states and their propagation through the detector \cite{Arvanitaki:2005nq, Kraan:2004tz,Mackeprang:2006gx,Mackeprang:2010}. 

There is thus some model dependence in estimating the number of stopped particles for a given $X$. To address this, we compute stopping fractions and give mass reaches with the conservative assumption that MMCPs are produced only by direct pair production. For colored MMCPs, hadronization fractions to different R-hadron states are as in \cite{Mackeprang:2010}.\footnote{It is possible that the charged R-hadrons decay strongly to a neutral state, depleting the population of charged states before they reach the ECAL. However, such large splittings in the R-hadron mass spectra are disfavored \cite{Foster:1998wu,Gates:1999ei}.}  We ignore the additional hadronic interactions of R-hadrons with the detector here as on average $\lesssim1$ such interaction will take place before the particle leaves the ECAL \cite{Mackeprang:2010} (in contrast, hadronic interactions are much more important for determing the total stopping rate in the HCAL).

To determine the stopping fractions in Table~\ref{tab:eventrate},  we have used MadGraph 5 to calculate the tree-level cross section for pair production of a MMCP traveling toward the center of the ECAL half barrel with $|\eta|<0.3$ and velocity less than the critical velocity $\beta$. For larger values of $\eta$, the physical size of the calorimeter cells is notably increased, and thus the resolution for measurements of stopped particle decays would decrease. We note however that a less conservative cut of $\eta < 1.3$ as used in late decay discovery searches \cite{Khachatryan:2010uf}  increases stopping fractions by a factor of 10-30\% depending on the MMCP properties.  For a MMCP of mass $M$, the critical velocity $\beta$ is determined by \cite{Arvanitaki:2005nq}
\begin{equation}
 x = x_0 \left(\frac{M}{500{\rm~GeV}}\right) \int_{\beta_0}^\beta \frac{{\beta'}^3 d\beta'}{1 + \log(\beta')/\kappa}.
\end{equation}
The ATLAS and CMS ECALs have roughly the stopping power of $x=20{\rm~ cm}$ of lead for a centrally produced X, and $x_0 = 503{\rm~m}$, $\kappa=3.6$, and $\beta_0\approx0.05$ for lead. For colored particles the final cross section has been scaled by higher order results for the total cross section.

\end{document}